\documentclass[11pt]{article}
\usepackage{epsfig,graphicx,amssymb,amsfonts}
\usepackage{latexsym}
\newcommand{\be}{\begin{equation}}
\newcommand{\ee}{\end{equation}}
\newcommand{\bea}{\begin{eqnarray}}
\newcommand{\eea}{\end{eqnarray}}

\usepackage{cite}
\usepackage{epsfig}
\usepackage{amsfonts}

\def \beq {\begin{equation}}
\def \eeq {\end{equation}}
\def \bea {\begin{eqnarray}}
\def \eea {\end{eqnarray}}
\def \nn {\nonumber}

 \def\e{{\rm e}}

\def\Z#1{_{\lower2pt\hbox{$\scriptstyle#1$}}}
\def\X#1{_{\lower2pt\hbox{$\scriptscriptstyle#1$}}}

\def\ApJ#1{Astrophys.\ J.\ {\bf#1}}

\begin{document}
\baselineskip=20pt

\vspace*{0cm}

\begin{flushright}
\end{flushright}

\begin{center}
{\Large{\bf A Note on Agegraphic Dark Energy}}

\vspace*{0.3in} Ishwaree P.\ Neupane \vspace*{0.3in}

\it Department of Physics and Astronomy, University of Canterbury\\
Private Bag 4800, Christchurch 8020, New Zealand\\
{\tt ishwaree.neupane@canterbury.ac.nz} \\
\vspace*{0.2in}
\end{center}

\begin{abstract}

Recently a new model of dynamical dark energy, or time-varying
$\Lambda$, was proposed by Cai [arXiv:0707.4049] by relating the
energy density of quantum fluctuations in a Minkowski space-time,
namely $\rho_q \equiv 3 n^2 m_P^2/t^2$, where $n\sim {\cal O}(1)$
and t is the cosmic time, to the present day dark energy density.
In this note, we show that the model can be adjusted to the
present values of dark energy density parameter $\Omega_q$
($\simeq 0.73$) and the equation of state ${\rm w}\Z{q}$ ($\simeq
-1$) only if the numerical coefficient $n$ takes a reasonably
large value ($n\gtrsim 3$) or the present value of the
gravitational coupling of q-field to (dark) matter is also
nonzero, namely, $\tilde{Q}\simeq
\frac{2}{n}(\Omega\Z{q0})^{3/2}>0$ where $\Omega\Z{q0}$ is the
present value of dark energy density fraction. We also discuss
some of the difficulties of this proposal as a viable dark energy
model with a constant $n$; especially, the bound imposed on the
dark energy density parameter $\Omega_q <0.1$ during big bang
nucleosynthesis (BBN) requires $n< 1/6$. To overcome this
drawback, we outline a few modifications where such constraints
can be weakened or relaxed. Finally, by establishing a
correspondence between the agegraphic dark energy scenario and the
standard scalar-field model, we also point out some interesting
features of an agegraphic quintessence model.

\end{abstract}

\vfill
\newpage
\baselineskip=20pt

\section{Introduction} Dark energy, or a mysterious force
propelling the universe, is one of the deepest mysteries in all of
science. This mysterious force now thought to account for about
73\% of the density of the entire universe~\cite{WMAP} came to
many's surprise in 1998, when the Supernova Cosmology Project and
the High-Z Supernova Search teams~\cite{supernovae} independently
announced their discovery that the expansion of the universe is
currently accelerating. One possible source of this late-time
cosmic acceleration is a form of energy known as the Einstein's
cosmological constant $\Lambda$ - a vacuum energy of empty space,
which acts like a perfect fluid with an equation of state ${\rm
w}\Z{\Lambda}=p\Z{\Lambda}/\rho\Z{\Lambda}=-1$.

In physics, it may be true that we do not have to go around a very
complicated (cosmological) model to explain the concurrent
universe. By somehow consistent with this idea, it has often been
argued by many that the mysterious dark energy we see today may
well be the manifestation of the Einstein's cosmological constant.
However, although appealing, this simplest explanation is in
blatant contradiction with all known calculations of zero-point
(vacuum) energy in quantum field theories~\cite{Weinberg:1988}. No
theoretical model, not even the most sophisticated, such as
supersymmetry or string theory~\cite{Green-etal}, is able to
explain the presence of a small positive cosmological constant in
the amount that our observations require~\cite{WMAP},
$\rho_\Lambda\sim 10^{-47}~{\rm GeV}^4$. If $\rho\Z{\Lambda}$ is
to be interpreted as the present-day dark energy density, then the
most pressing issue would be an understanding of why
$\rho\Z{\Lambda}^{1/4}$ is fifteen orders of magnitude smaller
than the electroweak scale ($M\Z{\rm EW}\sim 10^{12}~{\rm eV}$) -
the energy domain of major elementary particles in standard model
physics, and also why $\Omega\Z{\Lambda}\sim 3 \Omega\Z{\rm
matter}$ now.

Needless to say, that the most popular alternative to the
cosmological constant, which uses a dynamical scalar field $\phi$
with a suitably defined scalar field potential
$V(\phi)$~\cite{Wetterich:94,Zlatev:1998}, predicts a small (but
still an appreciable) deviation from the central prediction of
Einstein's cosmological constant, i.e. ${\rm w}\Z{DE}=-1$. Also,
the models of holographic dark energy~\cite{MLi:04} and agegraphic
dark energy~\cite{Cai:07a}, which both appear to be consistent
with quantum kinematics, in the sense that these models obey the
Heisenberg type uncertainty relation, predict a time-varying dark
energy equation of state, ${\rm w}\Z{\rm DE}>-1$. The cosmological
observations only suggest that ${\rm w}\Z{\rm DE} < -0.82$ (see,
for example, Ref.~\cite{Melchiorri02}). Clearly, there remains the
possibility that the gravitational vacuum energy is fundamentally
variable. In this Letter we discuss about this possibility in a
framework of the model of `agegraphic' dark energy recently
proposed by Cai~\cite{Cai:07a}. By adopting the viewpoint that the
standard scalar field models are effective theories of an
underlying theory of dark energy, we also establish a
correspondence between the agegraphic dark energy model and the
standard scalar field cosmology.

\section{Agegraphic dark energy}

Based on an intuitive idea developed by C. Mead in 1960s and its
generalization by K\'arolyh\'azy~\cite{Kaloryhazy:66a}, Ng and van
Dam~\cite{Ng-vanDam}, Maziashvili~\cite{Maziashvili:07a},
Sasakura~\cite{Sasakura} and others, Cai recently proposed a model
of dark energy, which he called agegraghic~\cite{Cai:07a}. In this
proposal, the present-day vacuum energy density is represented by
the energy density of metric fluctuations in a Minkowski
space-time
\begin{equation}\label{quantum-CC}
\rho\Z{q} \equiv \rho\Z{\Lambda} \propto \frac{1}{t^2\,l\Z{P}^2}
\sim \frac{m\Z{P}^2}{t^2} \equiv \frac{3 n^2\,m_{p}^2}{t^2},
\end{equation}
where the numerical coefficient $n \sim {\cal O}(1)$ and $l_{P}$
is Planck's scale. For the derivation of Eq.~(\ref{quantum-CC}),
we refer to the original
papers~\cite{Kaloryhazy:66a,Ng-vanDam,Maziashvili:07a,Sasakura}.
This idea per se is not totally new; many cosmological models
which involve discussion of a time-varying vacuum energy either
predict or demand similar scaling solutions. Although the
expression~(\ref{quantum-CC}) is based on a limit on the accuracy
of quantum measurements~\cite{Kaloryhazy:66a,Maziashvili:07a}, or
thought experiments, it can also be motivated by various field
theoretic arguments, see, e.g.~\cite{Wetterich:94,Lopez:1995}.
According
to~\cite{Kaloryhazy:66a,Ng-vanDam,Maziashvili:07a,Sasakura} the
total quantum fluctuations in the geometry of space-time can be
non-negligible (as compared to the critical mass-energy density of
the universe) when one measures them on long distances, like the
present linear size of our universe!

What may be particularly interesting in Cai's
discussion~\cite{Cai:07a} is that one may take the cosmic time
\begin{equation}\label{age-universe} t=\int_{0}^a
\frac{da}{H\,a}=\int H^{-1} d\ln a
\end{equation}
(up to an arbitrary constant) as the age of our universe, where
$a(t)$ is the scale factor of a Friedmann-Robertson-Walker
universe and $H \equiv \dot{a}/a$ is the Hubble parameter (the dot
denotes a derivative with respect to cosmic time $t$). This
implies ${d t}/{d\ln a}={1}/{H}$. Then, using the definition
\begin{equation}
\Omega\Z{q}\equiv \frac{\kappa^2\rho\Z{q}}{3 H^2}=\frac{n^2}{t^2
H^2},\label{def-Omega-q}
\end{equation}
(where $\kappa$ is the inverse Planck mass $m_{P}^{-1}=(8\pi
G_{N})^{1/2}$) and differentiating it with respect to e-folding
time ${\cal N}\equiv \ln a$, we get
\begin{equation}\label{constr-org}
\Omega\Z{q}^\prime +2 \varepsilon\Omega\Z{q} + \frac{2}{t H}
\Omega\Z{q} =0.
\end{equation}
where the prime denotes a derivative with respect to e-folding
time, $N =\ln a$, and $\varepsilon\equiv \frac{\dot{H}}{H^2}$.
Although $n$ can take either sign, we take $n>0$ and
$t=\frac{n}{H\sqrt{\Omega\Z{q}}}>0$. Hence
\begin{equation}\label{constr-old}
\Omega\Z{q}^\prime +2 \varepsilon\Omega\Z{q} + \frac{2}{n}
\left(\Omega\Z{q}\right)^{3/2} =0,
\end{equation}
Eq.~(\ref{constr-old}) may be supplemented by the conservation
equation for the field $q$:
\begin{equation}
\dot{\rho}\Z{q} + 3 H \rho\Z{q} \left(1 + {\rm w}_{q}\right) = 0,
\end{equation}
or, equivalently,
\begin{equation}
\Omega_{q}^\prime+ 2\varepsilon \Omega_{q} + 3(1+{\rm
w}_{q})\Omega_{q}=0.\label{agegraph-2}
\end{equation}
By comparing eqs.~(\ref{constr-old}) and (\ref{agegraph-2}) we get
\begin{equation}\label{main-relation}
{\rm w}\Z{q}= - 1 + \frac{2}{3\,n}\, \sqrt{\Omega\Z{q}}.
\end{equation}
This shows that the energy density $\rho\Z{q}$ emanating from the
space-time itself may act as a source of gravitational repulsion,
provided that $\sqrt{\Omega\Z{q}}<n$. This can be seen by
considering a pure de Sitter solution for which $3H^2=\rho\Z{q}$.
By inverting the relation (\ref{quantum-CC}), i.e. $t\equiv
\frac{n}{H\sqrt{\Omega\Z{q}}}$, and using
Eq.~(\ref{age-universe}), we find
\begin{equation}
a(t)=\left(\frac{c_1 t + c_2}{n}\right)^n, \end{equation} where
$c_1, c_2$ are arbitrary constants. For $n>1$, the $q$-field
behaves like a standard scalar field (or an inflaton), leading to
an accelerated expansion. However, this is just an ideal
situation; in practice $\Omega\Z{m}$ is never zero. Moreover,
since $\rho_q$ is decreasing with the cosmic time~\footnote{Note
that $t H=1/2$, $2/3$ and $>1$, respectively, during radiation,
matter and dark energy dominated phases.}, the ratio
$\rho\Z{m}/\rho\Z{q}$ could be relevant for all times! That is to
say, in the present universe, a small $n$ in a close proximity of
being unity cannot give an accelerated expansion. Specifically,
with the input $\Omega\Z{q}=0.73$, we get ${\rm w}\Z{q}< -0.82$
only for $n>3.16$. With such a large value of $n$, however, the
model cannot satisfy the bound $\Omega\Z{q} (1\,{\rm MeV})< 0.1$
imposed during the big bang nucleosynthesis (BBN) epoch unless one
modifies certain premises of the standard model cosmology (see
below). That is to say, the form of the so-called agegraphic dark
energy as presented in (\ref{quantum-CC}) is problematic, if the
present universe consists of only matter and this ``dark energy".
Does this mean that the model is already inconsistent with
observations? The answer is probably not. The present model may
yield some desirable cosmological features with some simple
modifications, such as
\begin{equation}\label{non-zero-t0}
\Omega\Z{q}\equiv \frac{n^2}{H^2 (t+\delta)^2},  \quad
t+\delta\equiv \frac{n}{H\sqrt{\Omega\Z{q}}}, \quad t\equiv
\int_0^a \frac{da}{aH}.
\end{equation}
The rescaling $t\to t+\delta$ does not affect the equations like
(\ref{constr-old}) and (\ref{main-relation}). For brevity, we
shall assume that $\delta\ge 0$ unless specifically specified.

Regardless of the choice of $\delta$, it is not sufficient to
concentrate only on the gravitational sector of the theory when
studying the concurrent cosmology. In order to study the
transition between deceleration and acceleration, one has to
consider the ordinary matter field, which is also the constituent
that we know dominated the universe in the past. To this end, one
supplements the evolution equation~(\ref{constr-old}) by
conservation equations for the ordinary fields (matter and
radiation). With the standard assumption that matter is
approximated by a non-relativistic pressureless fluid component
(${\rm w}_{m}\simeq 0$), and using the Friedmann constraint
$\Omega\Z{m}+\Omega\Z{q}+\Omega\Z{r}=1$, we find
\begin{eqnarray}\label{fractions}
\Omega\Z{q}&=& 1-\left(1+c\Z{0} \e^{\ln a}\right)\Omega\Z{r},
\quad \Omega\Z{m}=\Omega\Z{r} c\Z{0} \e^{\ln a},
\end{eqnarray}
where we have used the conservation equations
$\dot{\rho}\Z{m}+3H(1+{\rm w}\Z{m})\rho\Z{m}=0$ and
$\dot{\rho}\Z{r}+ 4 H\rho\Z{r}=0$. Thus, if an explicit functional
form of $\Omega\Z{q}$ is known, then $\Omega\Z{r}$ and
$\Omega\Z{m}$ can be known. The numerical coefficient $c\Z{0}$ in
Eq.~(\ref{fractions}) can be fixed using observational inputs:
ideally, $\Omega\Z{m 0}\simeq 0.27$ and $\Omega\Z{r}\simeq 5\times
10^{-5}$ at the present epoch ($a\simeq a_0 \equiv 1$) implies
that $c\Z{0}\simeq 5400$. For future use, we also define $\e^{\ln
a}=(1+z)^{-1}$, so that $a=a\Z{0}\equiv 1$ at $z=0$ ($a\Z{0}$ is
the present value of scale factor).

\begin{figure}[ht]
\begin{center}
\hskip-0.3cm
\epsfig{figure=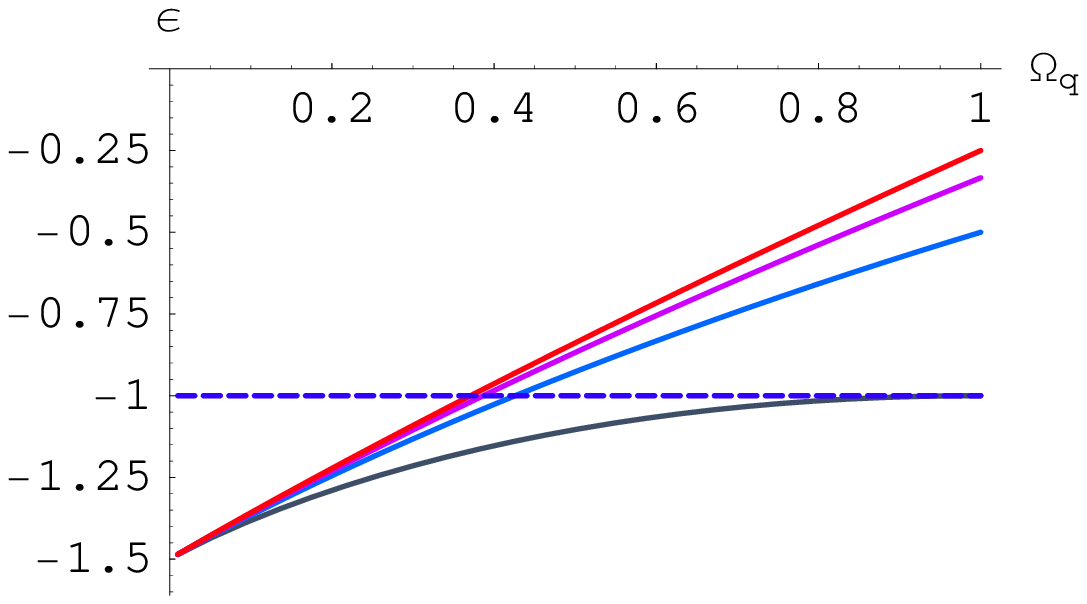,height=2.1in,width=2.9in}
\hskip0.6cm \epsfig{figure=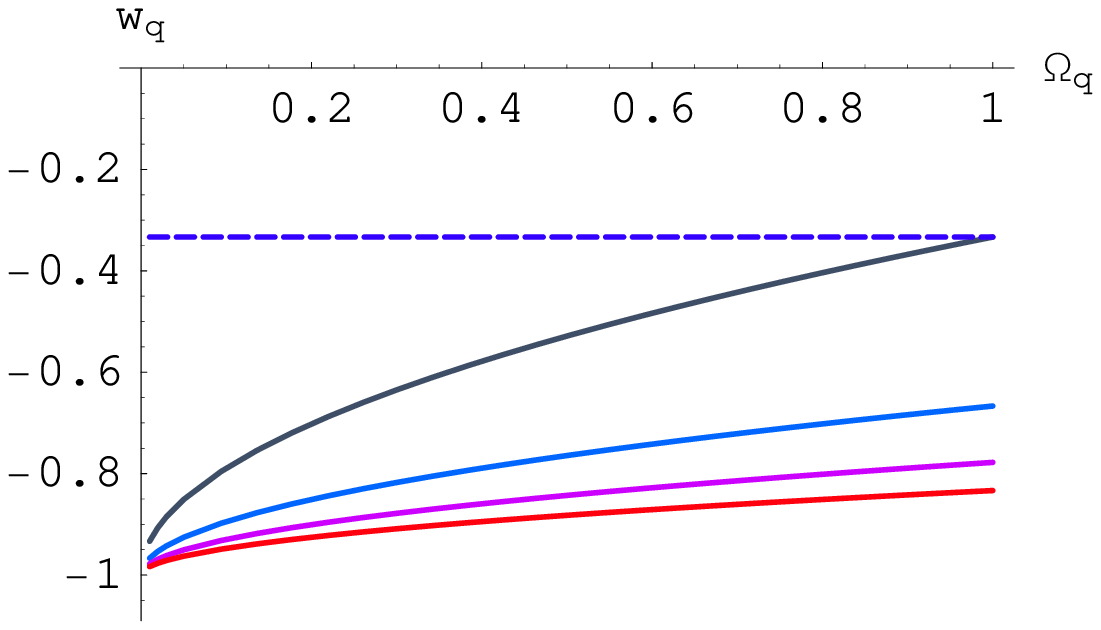,height=2.1in,width=2.9in}
\end{center}
\caption{The acceleration parameter $\varepsilon\equiv
\dot{H}/H^2$ (left plot) and the dark energy equation of state
${\rm w}\Z{q}$ (right plot) as functions $\Omega\Z{q}$, with
$n=1,2,3, 4$ (bottom to top (left plot) or top to bottom (right
plot)). Cosmic acceleration occurs for $\varepsilon>-1$ or ${\rm
w}_{\rm eff}<-1/3$. }\label{Fig1}
\end{figure}

All the examinations so far have been in a rather general way,
i.e. without making additional assumptions, except that ${\rm
w}\Z{m}\simeq 0$. For sure this is not really satisfying, as one
might be interested in analytic solutions of the system of
equations (\ref{constr-old}) and (\ref{fractions}). To this end,
we take $\Omega_{r}\approx 0$, which is also a reasonable
approximation valid at late times. From eqs.~(\ref{constr-old})
and (\ref{fractions}), we find
\begin{equation}\label{sol-epsilon}
\varepsilon=-\frac{1}{2}\frac{\Omega\Z{q}^\prime}{\Omega\Z{q}}
-\frac{1}{n}\sqrt{\Omega\Z{q}},
\quad \Omega_{m}+\Omega_{q}=1
\end{equation}
subject to the constraint
\begin{equation}\label{parametric-sol}
\ln a + C=
\frac{8}{3}\frac{\ln|3n-2\sqrt{\Omega_{q}}|}{(3n+2)(3n-2)}-
\frac{n \ln |1-\sqrt{\Omega_{q}}|}{3n-2}- \frac{n \ln
(\sqrt{\Omega_{q}}+1)}{3n+2}+ \frac{2}{3}\ln\sqrt{\Omega_{q}},
\end{equation}
where $C$ is an integration constant. Differentiating this last
equation with respect to $\ln a$, we get~\footnote{As already
noted in~\cite{Cai:07a,Cai:07b}), Eqs.~(\ref{constr-old}) and
(\ref{equality1}) hold not only for the form
$t=\frac{n}{H\sqrt{\Omega\Z{q}}}$, but also for $t
=\frac{n}{H\sqrt{\Omega\Z{q}}}+{\rm const}$.}
\begin{equation}
\Omega\Z{q}^\prime=(1-\Omega\Z{q}) (3-\frac{2}{n}
\sqrt{\Omega\Z{q}}) \Omega\Z{q}.\label{equality1}
\end{equation}
Substituting this expression back into Eq.~(\ref{sol-epsilon}), or
Eq.~(\ref{constr-old}), we find
\begin{equation}
\epsilon=-\frac{3n(1-\Omega\Z{q})-2\Omega\Z{q}^{3/2}}{2n}.
\end{equation}
This expression shows that the model can be consistent with
concordance cosmology, for which $\Omega\Z{q}\simeq 0.73$,
$\varepsilon>-1$ and $ {\rm w}\Z{q} <-0.82$, only if $n \gtrsim
3.16$. From the plots in Fig.~\ref{Fig1}, we can see that during
the matter dominated phase, $\Omega_m^\prime/\Omega_m\simeq {\rm
const}$~\footnote{The assumption that $a\sim 0$ in the matter
dominated phase was, however, not necessary, which led to an
apparent contradiction in~\cite{Cai:07a}. In the limit
$\Omega\Z{q}\to 0$, Eq.~(\ref{parametric-sol}) gives
$\Omega\Z{q}\propto a^3$~\cite{Cai:07a}. However, this solution
may not correspond to the matter dominated epoch; in any
consistent model, one should actually allow a nonzero
$\Omega\Z{r}$ in the limit $\Omega\Z{q}\to 0$. During matter
dominance one has $a\propto t^{2/3}$, $H^2=4/9 t^2$ and hence
$\Omega\Z{q}=9n^2/4(1+\delta/t)^2$. With $\delta\gtrsim {\cal
O}(10)\times t\Z{BBN}$, the present model could lead to some
desirable features even for $n\sim {\cal O}(1)$, thus the extra
parameter $\delta$ is a mixed blessing.}, $\epsilon\to -2/3$,
leading to $a\sim t^{2/3}$.

\begin{figure}[ht]
\begin{center}
\hskip-0.3cm \epsfig{figure=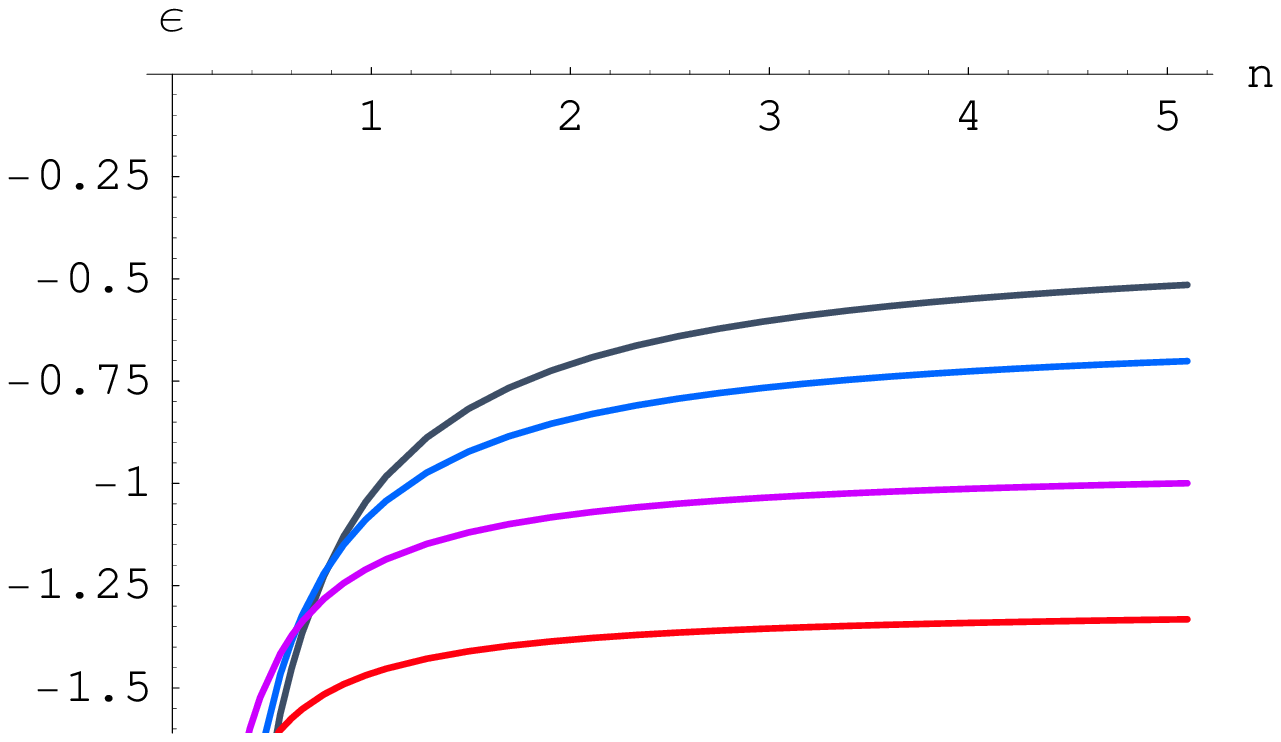,height=2.4in,width=3.2in}
\hskip0.2cm \epsfig{figure=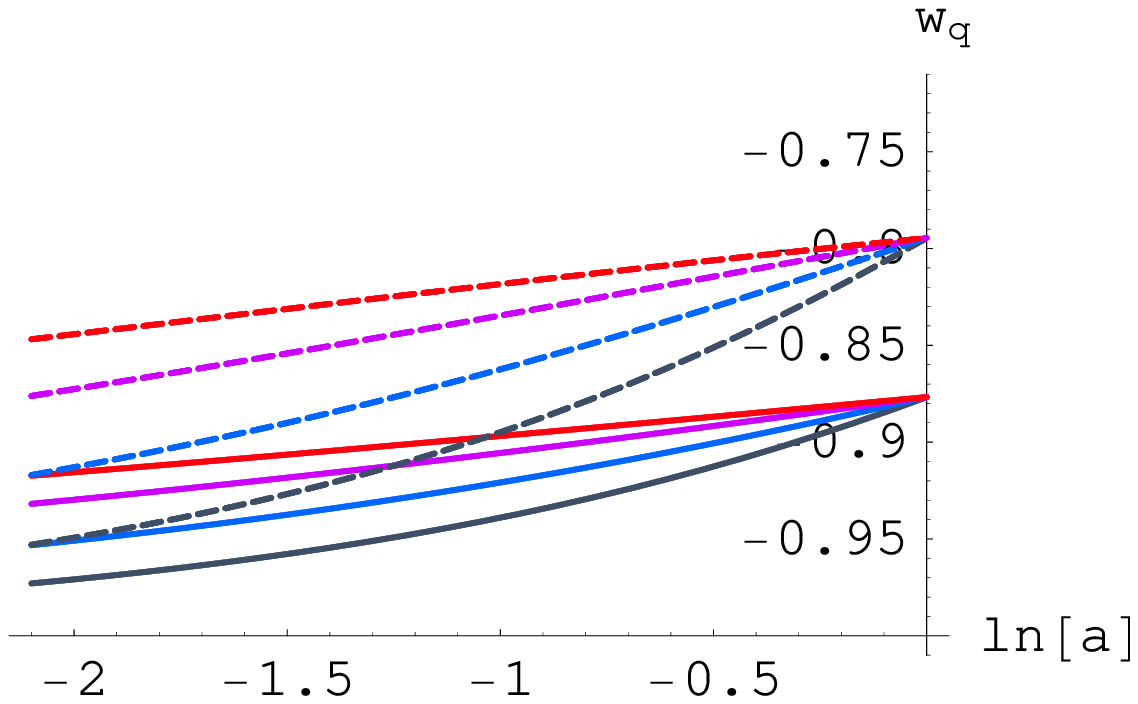,height=2.4in,width=3.2in}
\end{center}
\caption{(Left plot) $\epsilon$ as a function of $n$ with
$(\Omega\Z{q0},\Omega\Z{m0},\Omega\Z{r0})=(0.74,0.26,0),
(0.6,0.38,0.02), (0.4,0.5,0.1)$, $(0.3,0.6,0.2)$ (top to bottom).
(Right plot) ${\rm w}\Z{q}$ as a function of $\ln {a}$ with $n=3$
(dotted lines), $n=5$ (solid lines) and $m=2/3, 1, 3/2, 2 $
(bottom to top).}\label{Fig2}
\end{figure}

Next we study the system of equations with nonzero radiation
component. From equations (\ref{constr-old}) and
(\ref{fractions}), along with conservations equations
$\Omega_{m}^\prime+(2\epsilon+3)\Omega_{m}=0$ and
$\Omega_{r}^\prime+(2\epsilon+4)\Omega_{r}=0$, we get
\begin{equation}\label{sol-eps-RD}
\varepsilon=-\frac{1}{2}\frac{\Omega\Z{r}^\prime}{\Omega\Z{r}}-2,
\quad \frac{\Omega\Z{r}^\prime}{\Omega\Z{r}}=
\frac{3n\Omega\Z{m}-4n(1-\Omega\Z{r})+2\Omega\Z{q}^{3/2}}{n}.
\end{equation}
To solve the system of equations analytically, we need an extra
condition. Here we just want to check consistency of the model by
considering the following simplest solution~\footnote{This is a
fairly good approximation at a given epoch, such as $m\simeq 2/3$
during the matter-dominated epoch.}
\begin{equation}\label{power-law-sol-2}
a(t)= \left(c\Z{1} t+ c\Z{2}\right)^{m}, \quad
\sqrt{\Omega\Z{q}}=\frac{n}{m+n c\Z{0} a^{-1/m}}
\end{equation}
where $m$ is arbitrary. It should be emphasized that this solution
is valid for any value of $\delta$ in eqn.~(\ref{non-zero-t0}).
The integration constant $c\Z{0}$ may be fixed such that
$\Omega\Z{q}=\Omega\Z{q0}\simeq 0.73$ at $a=a\Z{0}=1$.
Fig.~\ref{Fig2} shows the behaviour of the acceleration parameter
$\epsilon$ and the dark energy equation of state ${\rm w}\Z{q}$.
With input $\Omega\Z{q 0}\simeq 0.73$, we clearly require $n > 3$
to get ${\rm w}\Z{q}<-0.82$ at the present epoch. This discussion
is consistent with the best-fit cosmological values of $n$ given
in Ref.~\cite{Myung07,Wu:07a}. Below we will consider the case of
{\it interacting} dark energy, for which the putative dark energy
field $q$ interacts non-minimally with (dark) matter.

\section{Interacting agegraphic dark energy}

In the non-minimal coupling case, the energy conservation
equations can be modified as
\begin{eqnarray}
0&=& \Omega_{q}^\prime+ 2\varepsilon \Omega_{q} +
3(1+{\rm w}_{q})\Omega_{q} + \widetilde{Q}, \label{agegraph4}\\
0&=& \Omega_{m}^\prime + 2\varepsilon \Omega_{m} + 3(1+{\rm
w}_{m})\Omega_{m}- \widetilde{Q},\label{agegraph5} \\
0&=& \Omega_r^\prime + 2\varepsilon \Omega_r +
4\Omega_r,\label{agegraph6}
\end{eqnarray}
where $\widetilde{Q}$ measures the strength of the gravitational
coupling of $q$-field to matter. In general, $\tilde{Q}=Q\Z{q}
\Omega\Z{m}$, where $Q\Z{q}\equiv \frac{d\ln A(q)}{d\ln a}$ and
$A(q)$ is a coupling function. In the minimal coupling or
noninteracting case $A(q)=1$. For simplicity, we will take ${\rm
w}\Z{m}\approx 0$ so that the matter is approximated by a
pressureless non-relativistic dust. Eqs. (\ref{constr-old}) and
(\ref{agegraph4})-(\ref{agegraph5}) can then be written as
\begin{eqnarray}
\epsilon=-\frac{1}{2}\frac{\Omega\Z{r}^\prime}{\Omega\Z{r}}-2,\nonumber
\\
\tilde{Q}=3\Omega\Z{m}-4(1-\Omega\Z{r})+\frac{2}{n}
\Omega\Z{q}^{3/2}-\frac{\Omega\Z{r}^\prime}{\Omega\Z{r}},\nonumber \\
w\Z{q}=-1+\frac{2}{3n}\sqrt{\Omega\Z{q}}-\frac{\tilde{Q}}{3\Omega\Z{q}}.
\end{eqnarray}
Although $\Omega\Z{r}\approx 0$ at the present epoch, the ratio
$\Omega\Z{r}^\prime/\Omega\Z{r}$ is non-negligible; in fact, the
value of $\Omega\Z{r}^\prime/\Omega\Z{r}$ should be less than $-2$
so as to allow an accelerated expansion ($\epsilon>-1$). Note
that, with $\tilde{Q}\ne 0$ the EoS parameter $w\Z{q}$ does not
explicitly depend on $n$ rather on the values of $\Omega\Z{q}$ and
$\Omega\Z{r}^\prime/\Omega\Z{r}$.

\begin{figure}[ht]
\begin{center}
\hskip-0.3cm \epsfig{figure=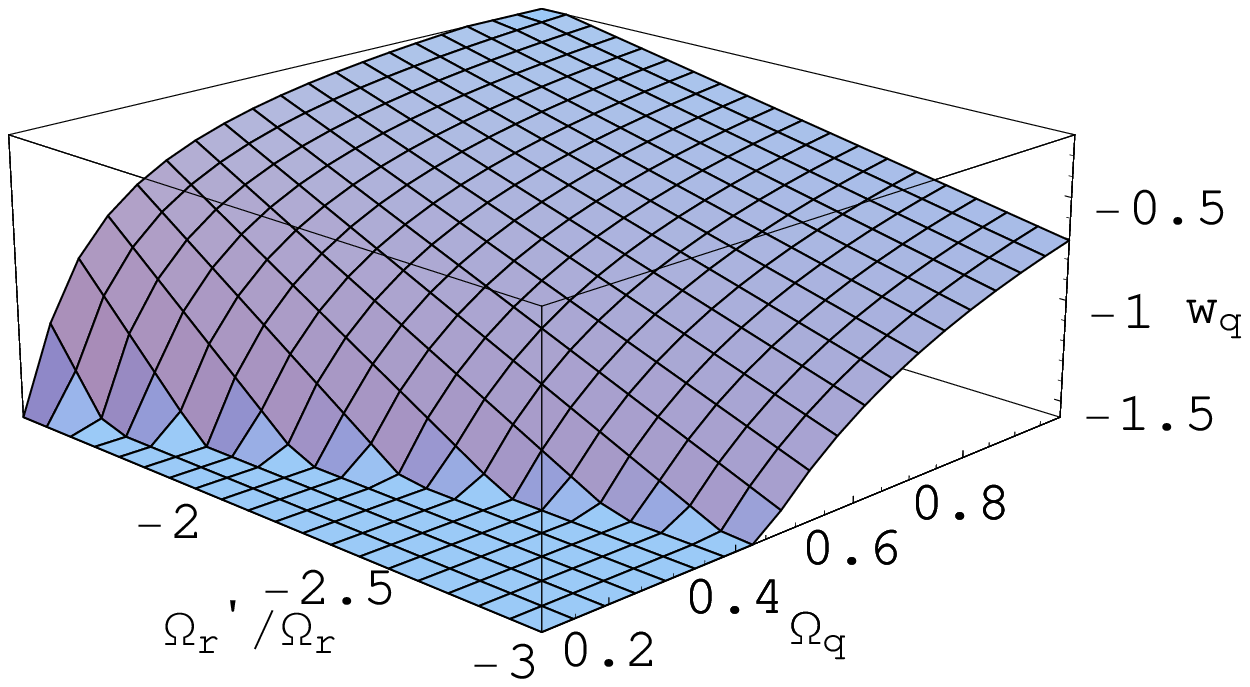,height=2.1in,width=2.8in}
\hskip0.4cm \epsfig{figure=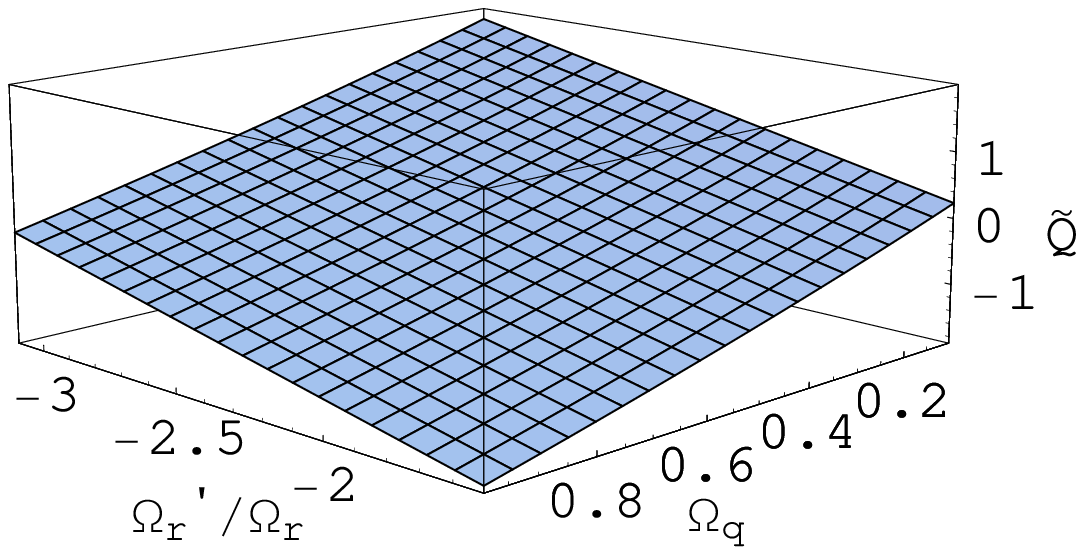,height=2.1in,width=2.8in}
\end{center}
\caption{The evolution of the EoS parameter ${\rm w}\Z{q}$ and the
coupling function $\tilde{Q}$ .}\label{Fig3}
\end{figure}

With $\tilde{Q}\ne 0$, one may get ${\rm w}\Z{q}\simeq -1 $ by
taking
\begin{equation}
\tilde{Q}_0\simeq \frac{2}{n} (\Omega\Z{q0})^{3/2},
\end{equation}
where $\Omega\Z{q 0}$ is the present value of $\Omega\Z{q}$. With
the input $\Omega\Z{q 0}=0.73$ and $n\sim {\cal O}(1)$, the
coupling $\tilde{Q}_0$ is relatively large. For this to happen,
the $q$-field should interact strongly at least with invisible or
dark matter~\footnote{It is a simplification when we say the
$q$-field couples to matter, when actually it is meant that
$q$-field couples to dark matter and that the baryonic component
is negligible. This discussion can easily be generalized to the
case where $\tilde{Q}_b=0$, in which case the $q$-field is coupled
only to dark matter, which then automatically satisfies possible
local gravity constraints.}. With further input that
$\Omega\Z{r}\simeq 0$ and $\Omega\Z{m 0}\simeq 0.27$, we get
$\Omega_r^\prime/\Omega_r =-3.19$ and hence $\epsilon=-0.405$,
which leads to an accelerated expansion, i.e. $a(t) \propto
t^{2.47}$.

The above discussion shows that in the case of a nontrivial
coupling between the $q$-field and matter, so that
$\widetilde{Q}\ne 0$, the model proposed in~\cite{Cai:07a} may be
adjusted to present-day dark energy parameters $\Omega_{q}\simeq
0.73$ and ${\rm w}_{q}\simeq -1$, if the present value of
$\widetilde{Q}$ is large, $\widetilde{Q}\sim {\cal O}(1)$ (see
Fig.\ref{Fig3}). However, this is not end of the story. As
mentioned above, with $\delta=0$ in (\ref{non-zero-t0}), the
present model finds stringent constraints in the early universe,
including the bound imposed on $\Omega\Z{q}$ during the BBN. To
quantify this, let us consider an epoch of cosmological expansion
where $t H\approx {\rm const}\equiv \alpha$. This then implies
that
\begin{equation}
\Omega\Z{q}\equiv \frac{\rho\Z{q}}{3 m\Z{P}^2 H^2}
=\frac{n^2}{\alpha^2},
\end{equation}
where we have used the relation~(\ref{quantum-CC}). The explicit
solution is then given by
\begin{eqnarray}
&& \Omega\Z{r}=\Omega\Z{r}^{(0)} \e^{-\,4\ln a} \e^{(2/\alpha)\ln
a},\quad \varepsilon=-\frac{1}{\alpha}, \quad
\Omega\Z{m}=1-\Omega\Z{q}-\Omega\Z{r},
\nonumber \\
&& \qquad \widetilde{Q}=\frac{2(2\alpha-1)}{\alpha} \Omega\Z{r} +
\frac{6(1+{\rm w})-2}{\alpha} \Omega\Z{m},
\end{eqnarray}
where ${\rm w}=0$ (${\rm w}=1/3$) for matter (radiation). During
the radiation dominance, one would expect that $a\propto t^{1/2}$,
implying that $\alpha= 1/2$ and thus $\Omega\Z{q} \simeq 4 n^2$.
If so, the above solution can satisfy the bound
$\Omega\Z{q}(1\,{\rm MeV}) < 0.1$ during BBN only if $n<1/6$,
indicating a small value of $n$ for which there would be no cosmic
acceleration at late times, satisfying $\Omega\Z{q 0}\simeq 0.73$
and ${\rm w}\Z{q}<-0.82$. For a consistent model cosmology,
perhaps one needs to satisfy during radiation-dominated epoch the
both conditions $\Omega\Z{q}\ll 1$ and $t H\simeq 1/2$,
simultaneously. Clearly, with $\delta=0$, the model of {\it
agegraphic} dark energy, which may be called {\it age-mapping},
cannot describe both the present and far past eras (including the
radiation-dominated universe) with a constant $n$, see also the
discussion in~\cite{Maziashvili:07a}. Nevertheless, as advertised
above, with some simple modifications the present model could lead
to a viable cosmological scenario. Let us in turn briefly discuss
them.

(1) A natural modification for which the numerical coefficient $n$
appearing in (\ref{quantum-CC}) varies slowly (actually,
increases) with time, such that $n(t\Z1)\ll n(t\Z2)$ where
$t\Z{2}\gg t\Z{1}$, could be compatible with concordance
cosmology, giving rise to standard conventional results, such as
$\Omega\Z 0\ll 1$ and $tH \simeq 1/2$ during the
radiation-domination epoch, and $\Omega\Z{q} \simeq 0.73$ and $
t\Z0 H\Z0 \simeq 1$ at the present epoch.

(2) Perhaps the most interesting possibility is to replace the
cosmic time $t$ by a conformal time $\eta$, as discussed recently
by Cai and Wei~\cite{Wei:2007ty}, and in more detail
in~\cite{Ish:07e}, for which $dt\equiv a d\eta$ and
\begin{equation}\label{new-wq-constr}
{\rm w}_{q}=-1+\frac{2}{3n}{\sqrt{\Omega_{q}}}\,e^{-\ln a}
-\frac{\widetilde{Q}}{3\Omega_{q}}.
\end{equation}
By setting $\widetilde{Q}=0$, and then comparing this equation
with the standard expression
\begin{equation}
{\rm w}\Z{q}=-1
-\frac{1}{3}\frac{\Omega\Z{q}^\prime}{\Omega\Z{q}}-\frac{2\varepsilon}{3},
\end{equation}
we get
\begin{equation}
\sqrt{\Omega\Z{q}}=\frac{n\e^{-\int\varepsilon \ln a}}{c+ \int
\e^{-\ln a} [\e^{-\int \varepsilon\, d\ln{a}}] \,d\ln{a}},
\end{equation}
where $c $ is an integration constant. This yields
\begin{equation}
\Omega\Z{q}= a^2 \left(\frac{1}{n} + \frac{c}{a}\right)^{-2} \quad
({\rm RD}),\qquad \Omega\Z{q}= a^2 \left(\frac{2}{n} +
\frac{c}{\sqrt{a}}\right)^{-2} \quad ({\rm MD}),
\end{equation}
respectively, for the radiation and matter dominated epochs. The
discussion in Ref.~\cite{Wei:2007ty} corresponds to the choice
$c=0$. Especially, in the case $\Omega\Z{q} \propto a^2$, the
limit $a\to 0$ can be regular, since ${\rm w}\Z{q}\to {\rm finite}
$ as $a\to 0$. The equation of state parameter ${\rm w}\Z{q}$
takes a finite value also in the early universe, provided that the
coupling term $\widetilde{Q}$ approaches zero faster than
$\Omega\Z{q}$.

(3) Another interesting possibility is to modify the expression
for $\rho\Z{q}$, Eq.~(\ref{quantum-CC}), itself, such that
\begin{equation}
\label{new-quan-CC} \rho\Z{q} \equiv \frac{3 n^2
m\Z{P}^2}{(t+\delta)^2},
\end{equation}
where now $\delta> 0$. This yields
\begin{equation}
\frac{n}{\sqrt{\Omega\Z{q}}}= t H \left(1+\frac{\delta}{t}\right).
\end{equation}
In the radiation-dominated universe $a(t)\propto t^{1/2}$ and
hence $H t\simeq 1/2$. Now, the BBN bound $\Omega\Z{q} (1~{\rm
MeV})<0.1$ can be satisfied by choosing $\delta$ such that $40 n^2
< (1+\delta/t)^2$. As a typical example, let us take $n= 3$, then
the BBN bound $\Omega\Z{q} (1~{\rm MeV})\lesssim 0.1$ is satisfied
for $\delta \gtrsim 18 \times t\Z{\rm BBN}$. Although the choice
$\delta=0$, being the most canonical, allows one to solve the
field equations analytically, the consistency of the model with
concordance cosmological requires $\delta>0$.

One may reconstruct an explicit observationally acceptable model
of evolution from the big bang nucleosynthesis to the present
epoch, by considering a general exponential
potential~\cite{Ish:06c}
$$ V(\phi)=V\Z{0} \exp\left(-\lambda \phi/m\Z{P}\right)$$ where
$\lambda\equiv \lambda(\phi)$. In the present model, this again
translates to the condition that the numerical coefficient $n$
(appearing in Eq.~(\ref{quantum-CC})) also becomes a slowly
varying function of cosmic time $t$ (or the age of the universe).
An explicit construction of such a model is beyond the scope of
this Letter.

\section{Agegraphic quintessence}

The {\it agegraphic} dark energy model discussed above can be
analysed also by considering the standard scalar field plus matter
Lagrangians
\begin{equation}\label{action} {\cal L} = \sqrt{-g} \left(
\frac{R}{2\kappa^2} -\frac{1}{2} (\partial \phi)^2- V(\phi)
\right)+{\cal L}_m.
\end{equation}
Without loss of generality, we will relate the putative dark
energy field $q$ (appearing in Eq.~(\ref{quantum-CC})) with the
standard scalar field $\phi$ by defining $\phi\equiv \phi(q)$. For
simplicity, let us first drop the matter part of the Lagrangian,
which will be considered later anyway. With the standard flat,
homogeneous FRW metric: $ds^2=-dt^2+ a^2(t) d{\bf x}^2$, we find
that the two independent equations of motion following from
Eq.~(\ref{action}) are given by
\begin{eqnarray}
 2 \dot{H}+ \kappa^2 \dot{\phi}^2 &=&0, \label{minimal1}\\
 \ddot{\phi}  +
3H\dot{\phi} &=&  -\frac{dV(\phi)}{d\phi}. \label{minimal2}
\end{eqnarray}
Eq. (\ref{minimal2}) can be written as
\begin{equation}
\dot{\rho}_{\phi} + 3 H \rho_{\phi} \left(1 + {\rm
w}_{\phi}\right) = 0,\label{minimal3}
\end{equation}
where ${\rm w}_{\phi}\equiv p_{\phi}/\rho_{\phi}$ and $\rho_{\phi}
\equiv \frac{1}{2} \dot{\phi}^2 + V(\phi)$. Using the definitions
\begin{equation}
\varepsilon\equiv \frac{\dot{H}}{H^2}, \quad
\Omega_{\phi}=\kappa^2\frac{\rho_{\phi}}{3H^2},
\end{equation}
we arrive at
\begin{eqnarray}
0&=& \Omega_{\phi}^\prime+ 2\varepsilon \Omega_{\phi} +
3(1+{\rm w}_{\phi})\Omega_{\phi} ,\label{new-eq1} \\
0&=&3{\rm w}_{\phi} \Omega_{\phi}+2\varepsilon+3.\label{new-eq2}
\end{eqnarray}
These equations may be solved analytically only by imposing one
extra condition, since the number of degrees of
freedom~\footnote{Here $a(t)$, $\phi(t)$ and $V(\phi)$ are primary
variables, while $\Omega\Z{\phi}$, $\varepsilon$ and ${\rm
w}\Z{\phi}$ are secondary (derived) variables.} exceeds the number
of independent equations.

For completeness, we write down the equations of motion by
considering the case where the putative dark energy field $\phi$
interacts with ordinary matter. The set of equations
(\ref{new-eq1})-(\ref{new-eq2}) are then modified as (see Appendix
for the details)
\begin{eqnarray}
0&=& \Omega_{\phi}^\prime+ 2\varepsilon \Omega_{\phi} +
3(1+{\rm w}_{\phi})\Omega_{\phi} + \widetilde{Q},\label{EoM1} \\
0&=& \Omega_{m}^\prime + 2\varepsilon \Omega_{m} +
3(1+{\rm w}_{m})\Omega_{m}- \widetilde{Q},\label{EoM2}\\
0&=&\Omega_{r}+ 3{\rm w}_{\phi} \Omega_{\phi}+ 3{\rm w}_{m}
\Omega_{m} +2\varepsilon+3.\label{EoM3}
\end{eqnarray}
Here $\widetilde{Q}$ measures the strength of a gravitational
coupling of $\phi$-field to matter. Without any restriction on
$\Omega_{\phi}$, or the potential $V(\phi)$, we find that the dark
energy EoS ${\rm w}_\phi$ is given by
\begin{equation}\label{de-EoS}
 {\rm w}_{\phi} = - \frac{2\varepsilon+3+ 3\sum_i
{\rm w}_{i}\Omega_{i}+\Omega_{r}}{3\Omega_{\phi}},
\end{equation}
where $i=m$ (matter) includes all forms of matter fields, such as
pressureless dust (${\rm w}=0$), stiff fluid (${\rm w}=1$) and
cosmic strings (${\rm w}=-1/3$). Note that the universe
accelerates when the effective equation of state ${\rm w}_{\rm
eff}$ becomes less than $-1/3$ (where ${\rm w}\Z{\rm eff} \equiv
-1-2\varepsilon/3$), not when ${\rm w}_{q}<-1/3$; it is because,
for a cosmic acceleration to occur, a gravitationally repulsive
force or dark energy must overcome a gravitational attraction
caused by ordinary matter and radiation.

\begin{figure}[ht]
\begin{center}
\hskip-0.4cm
\epsfig{figure=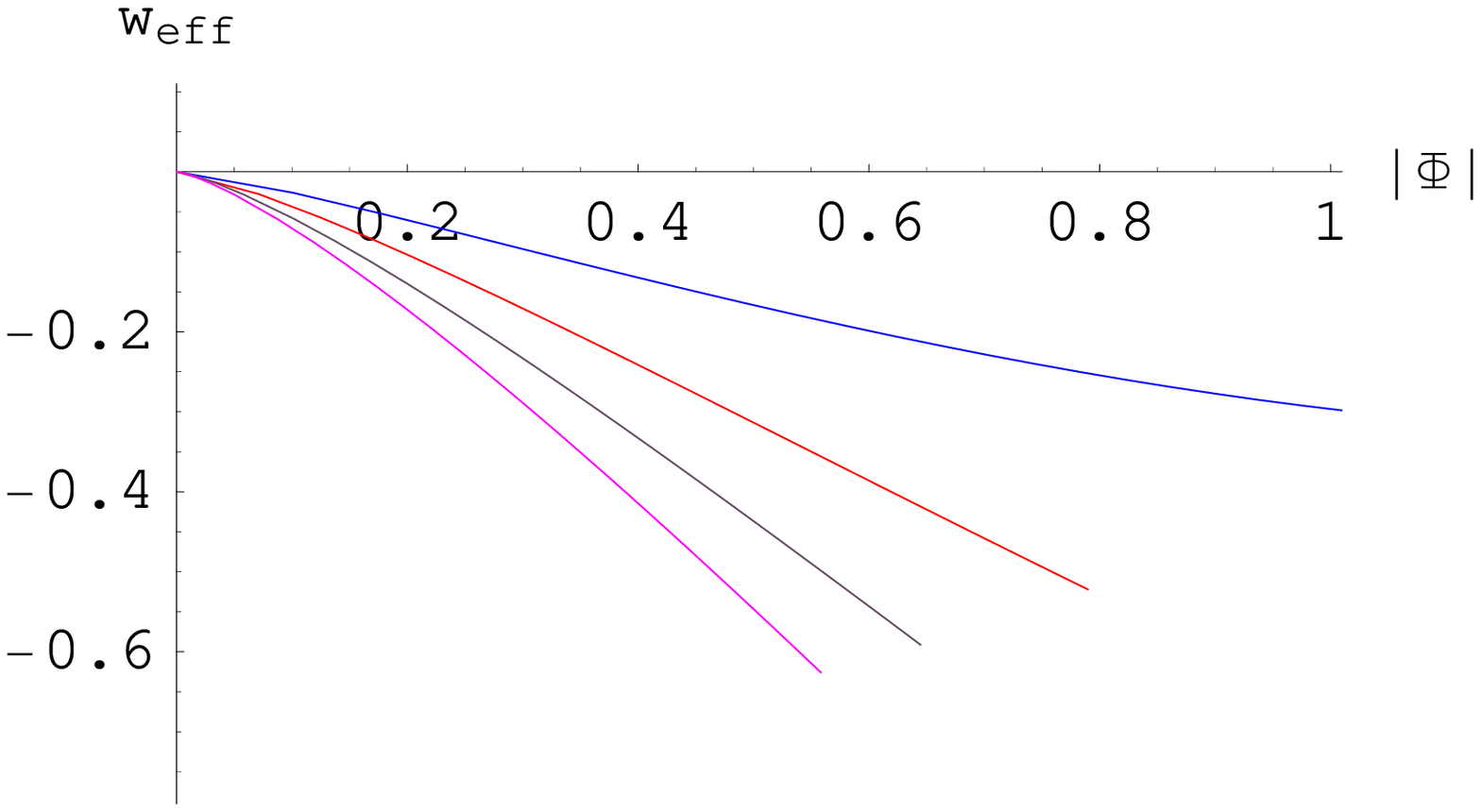,height=2.1in,width=3.1in}
\hskip0.1cm
\epsfig{figure=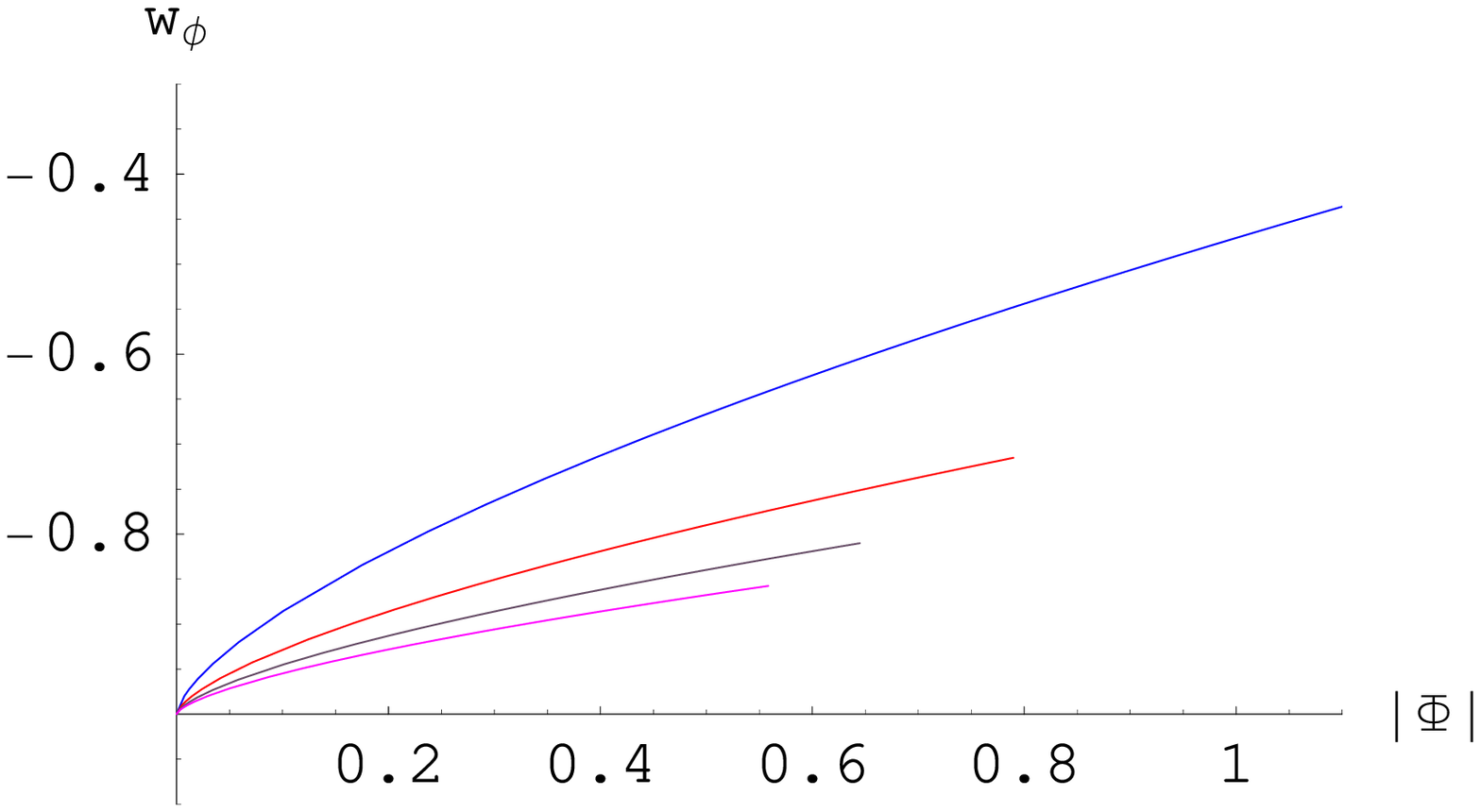,height=2.1in,width=3.1in}
\end{center}
\caption{The effective equation of state ${\rm w}\Z{\rm
eff}$($\equiv -1-2\varepsilon/3$) and dark energy equation of
state ${\rm w}\Z{\rm DE}\equiv {\rm w}\Z{\phi}$ as a function of
$\Phi$, with $n=1, 2, 3, 4$ (from top to bottom) and
$\Omega_\phi=(0, 0.73)$. The end point of each curve (or line)
corresponds to the value $\Omega\Z{\phi}=0.73$, while
$\Omega\Z{\phi}=0$ at $|\Phi|=0$. }\label{Fig4}
\end{figure}

In the particular case that ${\rm w}_{i}\approx 0$ and
$\Omega_{r}\approx 0$, the universe accelerates when ${\rm
w}\Z{\phi}\Omega\Z{\phi}<-1/3$, or when $\varepsilon>-1$, where
\begin{equation}\label{def-ep}
\varepsilon = -\frac{3(1+{\rm w}\Z{\phi}\Omega\Z{\phi})}{2}.
\label{reduced-ep}
\end{equation}
From the relations given below Eq.~(\ref{minimal3}), we can easily
derive
\begin{equation}\label{quin-relation}
\Phi^2 \equiv \frac{\dot{\phi}^2}{m\Z{P}^2
H^2}=3\Omega\Z{\phi}\left(1+{\rm w}\Z{\phi}\right), \quad U \equiv
\frac{V(\phi)}{m\Z{P}^2 H^2}=\frac{3}{2} \Omega\Z{\phi}
\left(1-{\rm w}\Z{\phi}\right).
\end{equation}
In order to reconstruct a model of agegraphic quintessence, one
may supplement these relations by the EoS of agegraphic dark
energy, ${\rm w}\Z{\phi}=-1+(2/3n)\sqrt{\Omega\Z{\phi}}$. From
Eq.~(\ref{def-ep}), we then find
\begin{equation}\label{epsilon-new}
\varepsilon\equiv \frac{\dot{H}}{H^2} =-\frac{3}{2}
(1-\Omega\Z{\phi})-\frac{\Omega\Z{\phi}^{3/2}}{n}.
\end{equation}
As expected, this expression of $\varepsilon$ matches with that
obtained from eqs.~(\ref{sol-epsilon}) and (\ref{equality1}). In
Fig.~\ref{Fig4} we show the behaviour of ${\rm w}\Z{\rm eff}$ and
${\rm w}\Z{\phi}$ with respect to a dimensionless parameter,
$\Phi$ ($\equiv |\dot{\phi}|/(m\Z{P} H)$). The plots there show
that the universe can accelerate (${\rm w}\Z{\rm eff}< -1/3$) only
if $n\gtrsim 2$, and $\Omega\Z{\phi}$ may evolve from zero to
higher values as the $\phi$-field starts to roll. The $\phi$-field
is almost frozen, i.e. $\dot{\phi}\simeq 0$, during the
matter-dominated phase where ${\rm w}\Z{\rm eff}\simeq 0$ or
$\epsilon\simeq -3/2$, while $\dot{\phi}$ is nonzero during an
accelerating (or dark energy dominated) regime, leading to
$U(\phi)\equiv V(\phi)/(m\Z{P}^2 H^2)>0$ at present.

\begin{figure}[ht]
\begin{center}
\hskip-0.4cm
\epsfig{figure=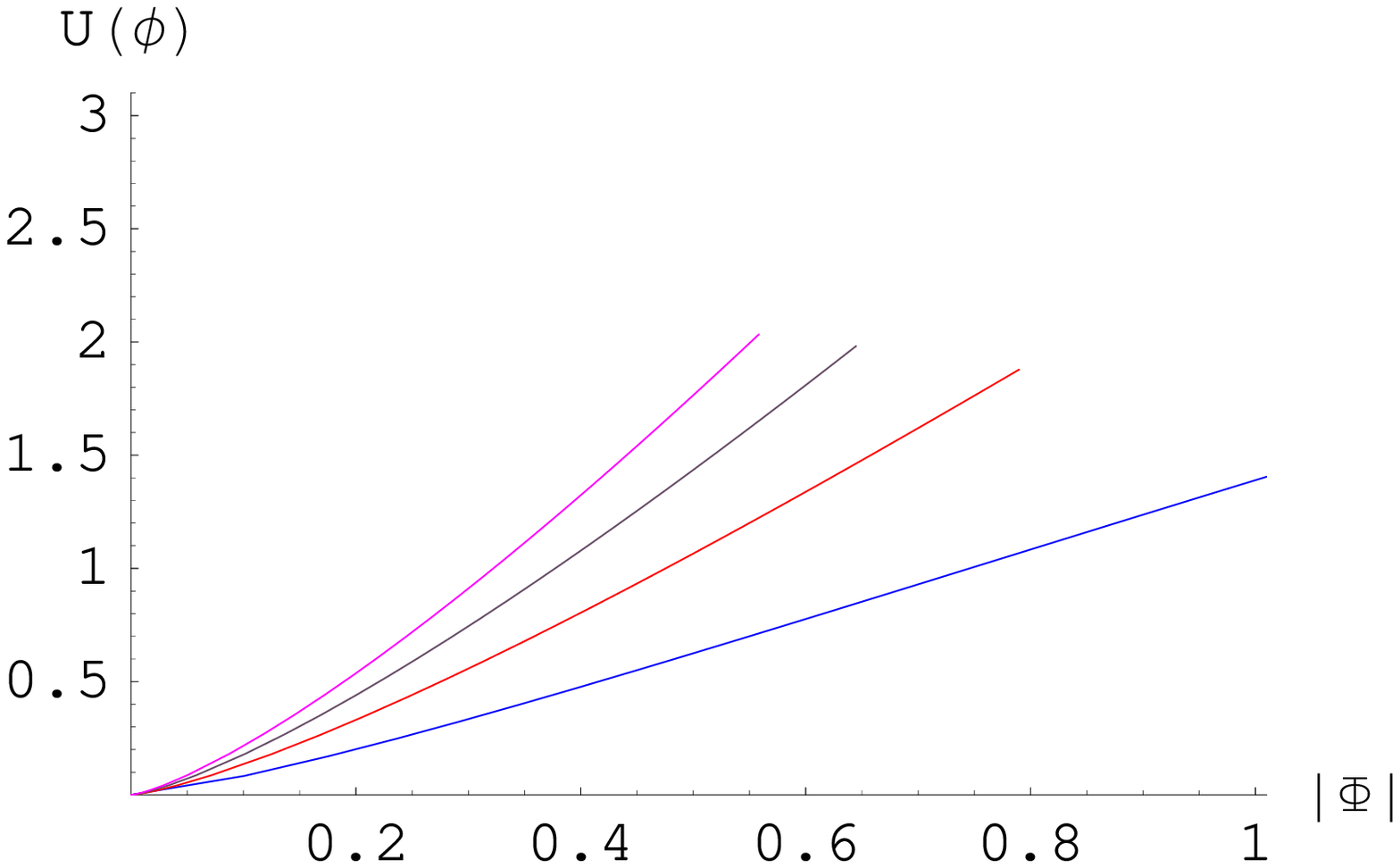,height=2.3in,width=3.2in}
\hskip-0.2cm
\epsfig{figure=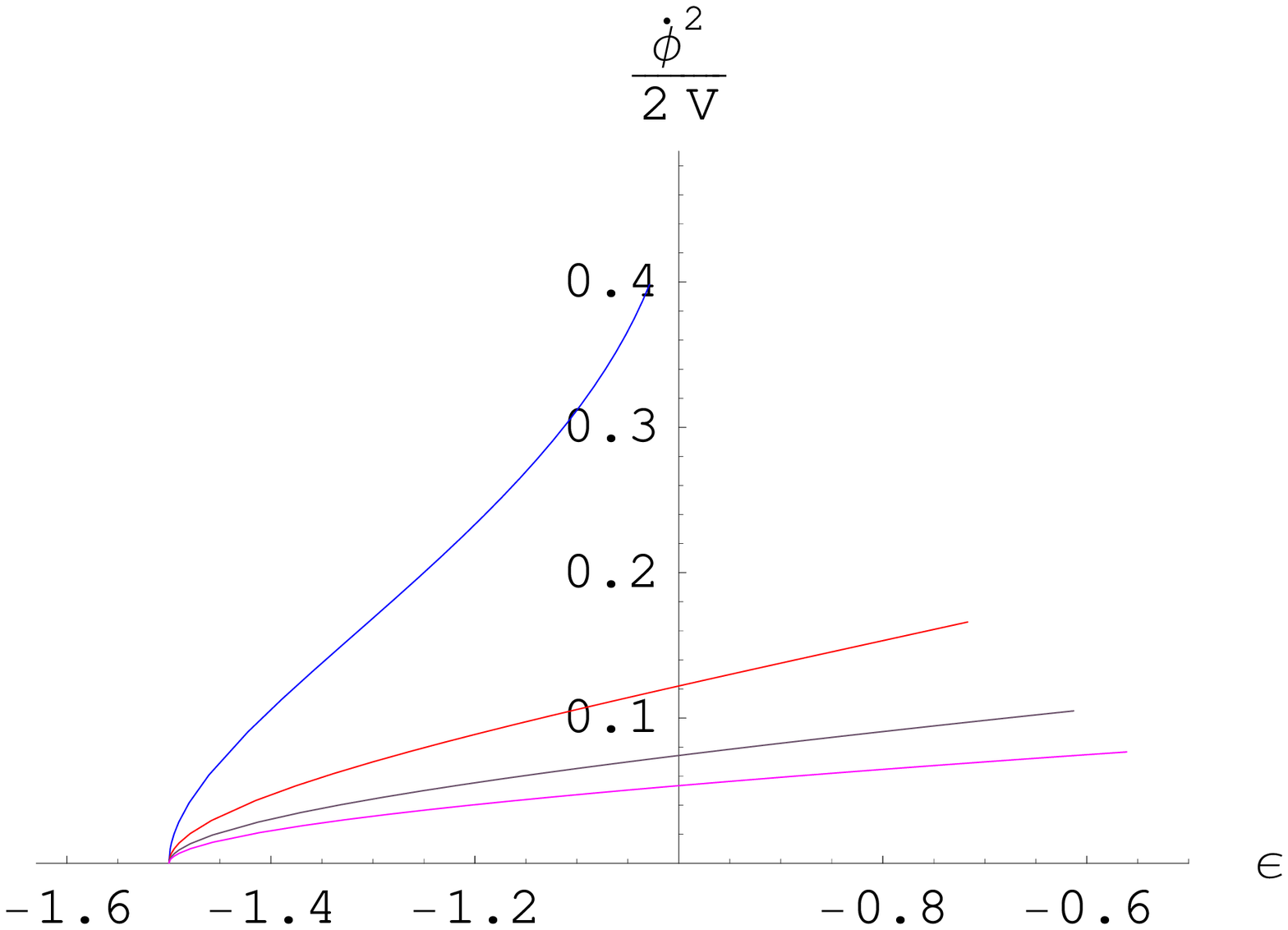,height=2.3in,width=3.4in}
\end{center}
\caption{(Left plot) Evolution of a normalised agegraphic
potential $U(\phi)$ with respect to $\Phi$, for $n=1, 2, 3, 4$
(bottom to top). (Right plot) The ratio $r\equiv \dot{\phi}^2/2V$
with respect to $\epsilon$, for $n=1, 2, 3, 4$ (top to bottom),
which usually measures the value of $(1+{\rm w}_\phi)/(1-{\rm
w}_\phi)$. Acceleration occurs when $\varepsilon>-1$. The end
point of each curve corresponds to $\Omega\Z{\phi}=0.73$. The
potential $V(\phi)$ vanishes at $|\Phi|=0$ (where
$\Omega\Z{\phi}=0$), while it increases as the density parameter
$\Omega\Z{\phi}$ grows. }\label{Fig5}
\end{figure}
\begin{figure}[ht]
\begin{center}
\hskip-0.4cm
\epsfig{figure=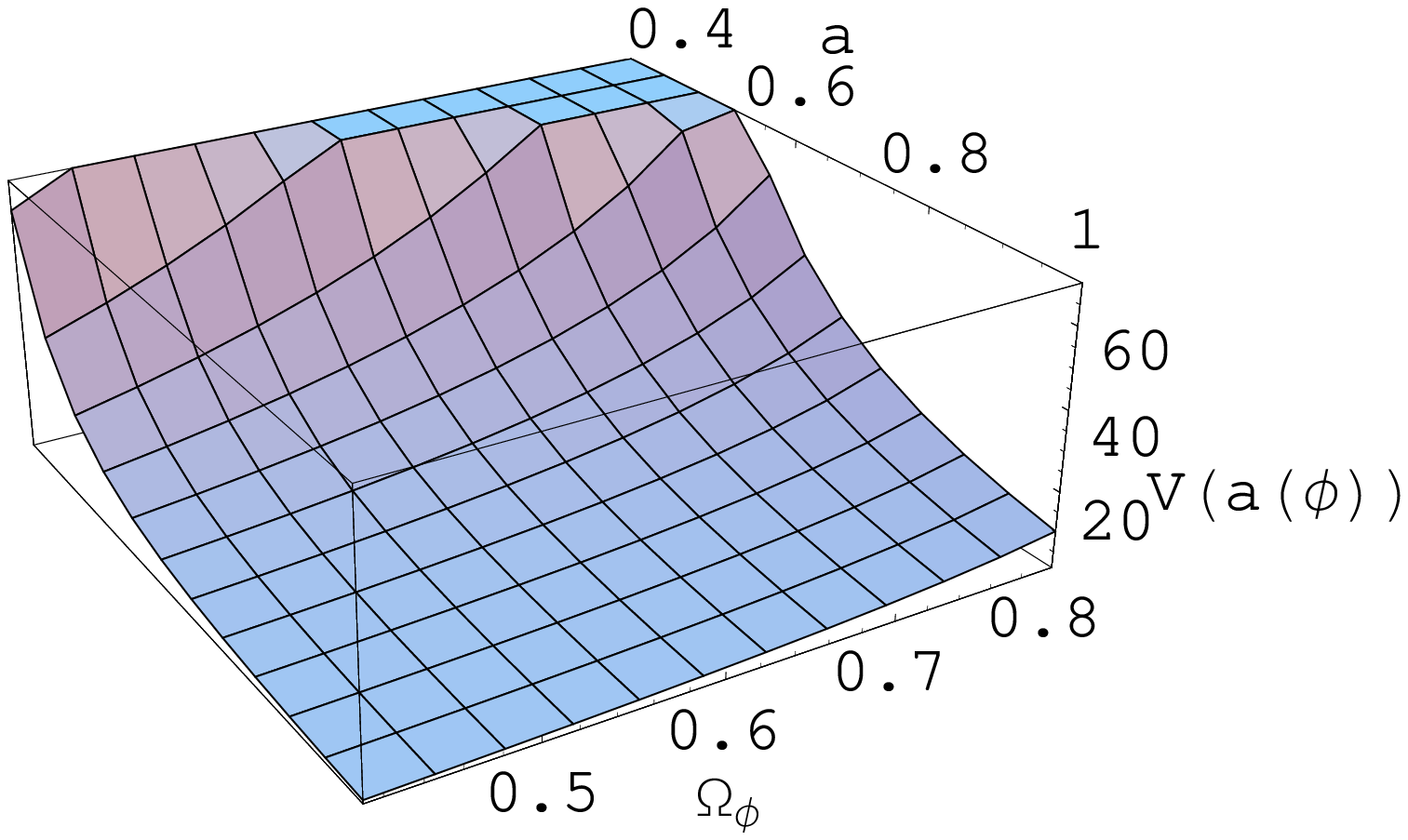,height=2.3in,width=3.5in}
\hskip-0.2cm
\epsfig{figure=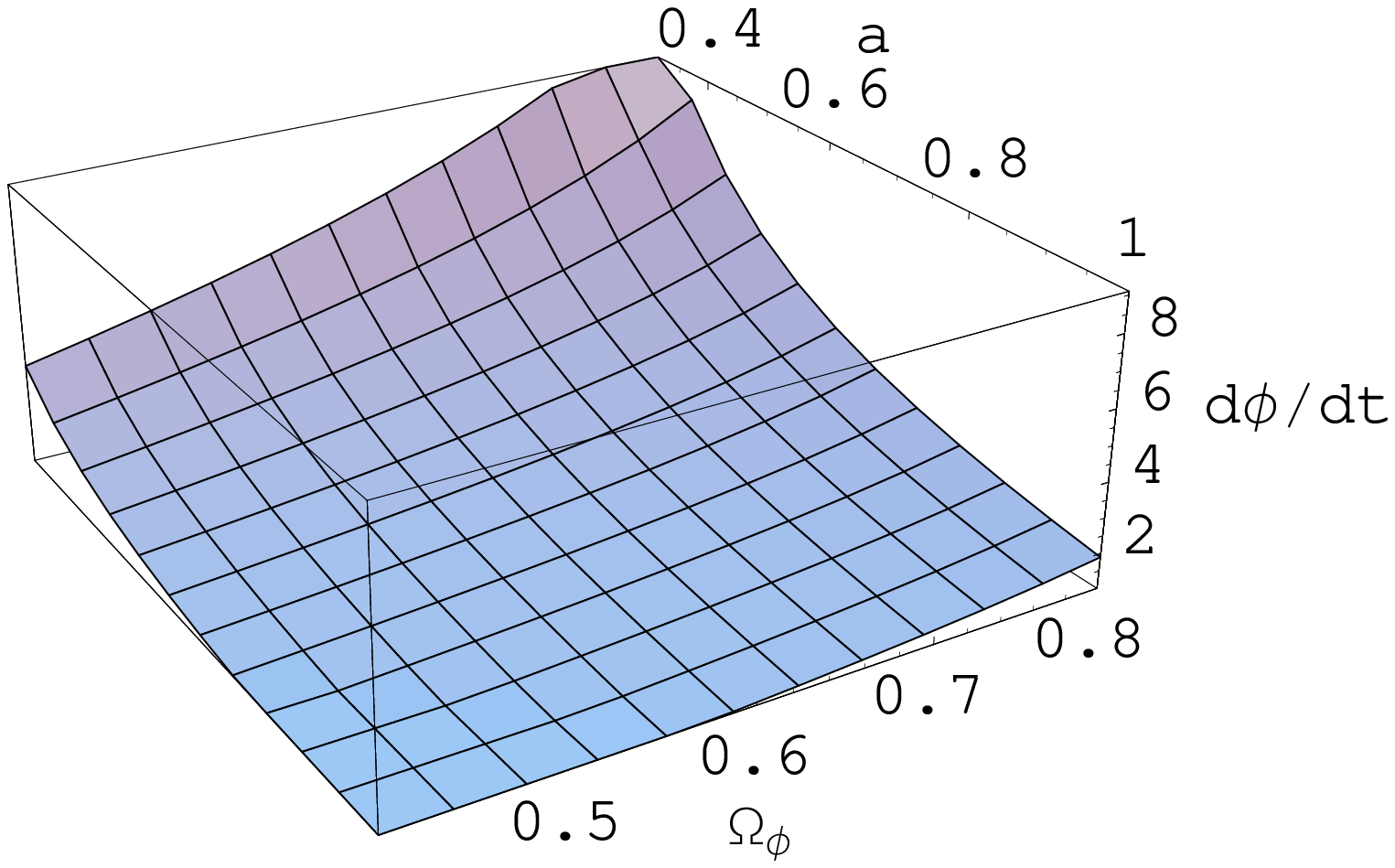,height=2.3in,width=3.2in}
\end{center}
\caption{The reconstructed potential $V(\phi)$ and the
time-derivative of $\phi$ as functions of $\Omega_\phi$ and scale
factor $a$. We have set $\sqrt{\Omega_{m0}}\, m\Z{P} H\Z{0}
=1$.}\label{Fig6}
\end{figure}

Fig.~\ref{Fig5} shows that the normalised agegraphic potential
$U(\phi)$ vanishes at $\Phi=0$. This feature is clearly different
from that of the standard quintessence model, for which,
generally, $V(\phi)={\rm const}$ at $\dot{\phi}=0$. Another
crucial difference is that as the universe evolves from a
matter-dominated epoch ($\varepsilon\simeq -3/2$) towards a dark
energy dominated epoch ($\varepsilon>-1$), the ratio
$\dot{\phi}^2/V(\phi)$ increases with respect to dark energy
density fraction $\Omega\Z{\phi}$, as well as with $\varepsilon$,
implying that the agegraphic quintessence model constructed above
falls into the `thawing' model~\cite{Caldwell05}, rather than the
`freezing' model for which $\dot{\phi}=0$ corresponds to an
analytic minimum of the potential. This behaviour is seen also
from the ratio $\dot{\phi}/H\propto \Omega_{\phi}^{3/2}$, which
increases as $\Omega_\phi$ increases.

To evaluate $V(\phi)$ we also need an analytic expression of
$H(a(\phi))$. From the Friedmann constraint,
$\Omega_\phi+\Omega_m=1$, we obtain
\begin{eqnarray}
&& 1-\Omega_\phi = \frac{\rho_m}{3H^2 m\Z{P}^2}=\frac{\rho_{m
0}}{3 H\Z{0}^2 m\Z{P}^2} \frac{H\Z{0}^2}{a^3 H^2}\equiv
\frac{\Omega_{m 0}}{a^3} \frac{H\Z{0}^2}{H^2}
\nn \\
\Rightarrow && H(a)=H\Z{0} \left(\frac{\Omega_{m
0}}{(1-\Omega_\phi) a^3}\right)^{1/2},
\end{eqnarray}
where for $1>\Omega_\phi>0$. From eqs.~(\ref{quin-relation}), we
then find
\begin{equation}
V(\phi)=\frac{3}{2} \Omega_{m 0} \, m\Z{P}^2 H\Z{0}^2
\sqrt{\Omega_{m 0}}\,\frac{\Omega_\phi (1-{\rm
w}_\phi)}{(1-\Omega_\phi) a^3},\quad \dot{\phi}=\sqrt{\Omega_{m
0}}\,m\Z{P} H\Z{0} \sqrt{\frac{3\Omega_\phi (1+{\rm
w}_\phi)}{(1-\Omega_\phi)a^3}}.
\end{equation}
We plot these quantities in Fig.~\ref{Fig6}. The left plot in
Fig.~\ref{Fig6} shows that $\Omega_\phi$ tends to increase the
potential while a growth in scale factor tends to decrease it.
Using the relation $\dot{\phi}/H\propto \Omega_\phi^{3/2}$, we
find that the potential is a slowly increasing exponential
function of $\phi$. Thus it is not surprising that the agegraphic
quintessence model draws a parallel with the simplest solution of
an exponential potential $V(\phi)\propto e^{-\sqrt{2}\, \lambda
(\phi/m\Z{P})}$, i.e. $\phi/m\Z{P}= (\sqrt{2}/\lambda) \ln
(t+t\Z{1})$ and $\rho\Z{\phi}\equiv
\frac{1}{2}\,\dot{\phi}^2+V(\phi) \propto m\Z{P}^2/(t+t\Z{1})^2$.

In the non-minimal coupling case, the energy conservation
equations can be modified as
\begin{equation}\label{main-cons-eqs}
\frac{d\rho_m}{da}+\frac{3}{a}\rho_m = +\widehat{\alpha} \rho_m,
\quad \frac{d\rho_\phi}{da}+\frac{3}{a}\rho_\phi(1+{\rm w}_\phi)
=-\widehat{\alpha} \rho_m,
\end{equation}
where $\hat{\alpha} \equiv - Q \beta(\phi) \frac{d(\kappa
\phi)}{da}=-\frac{d\beta}{d\phi} \frac{d\phi}{da}$ and $Q\equiv
d\ln \beta(\phi)/d(\kappa \phi)$ (see Appendix A) and we have
taken ${\rm w}_m=0$. The local gravity experiments provide some
constrains on the value of $Q$~\cite{Damour:1993id}~\footnote{More
precisely, $|Q| <0.1778$ or $Q^2=(1-\hat{\gamma})/(1+\hat{\gamma})
<0.0313$, or equivalently $|1-\hat{\gamma}|<2\times 10^{-3}$,
where $\hat{\gamma}$ is the PPN parameter.}, and, presumably, also
on $\hat{\alpha}$. In the particular case that
$\widehat{\alpha}\simeq {\rm const}$, or $\beta(\phi)\propto
a(\phi)+\beta\Z{0}$, we get
\begin{equation}
\rho_m(a)=\frac{\rho\Z{m 0}}{a^3}\,e^{-a \widehat{\alpha}},\quad
\rho_{\phi}=\left(\rho\Z{\phi 0}-\widehat{\alpha} \int \rho_m
\,\exp\left[3\int \frac{1+{\rm w}_\phi(a)}{a}\,da\right]da\right)
\exp\left[-3\int \frac{1+{\rm w}_\phi(a)}{a}\,da\right].
\end{equation}
The above two equations can be inverted to give
\begin{equation}\label{w-phi-general}
{\rm w}_\phi(a)=-1- \frac{a}{3}
\frac{d\ln\rho_\phi}{da}-\frac{\widehat{\alpha} \rho_m
a}{3\rho_\phi}.
\end{equation}
It is interesting to note that, for $\widehat{\alpha}
>0$, the dark energy equation of state becomes more negative as
compared to the $\widehat{\alpha} =0$ case. It is also plausible
that ${\rm w}_\phi(a)<-1$, if $\widehat{\alpha}\gtrsim {\cal
O}(1)$ is allowed.

Although it may not be essential, one can modify the conservation
equations, for example, as
\begin{equation}
\frac{d\rho_m}{da}+\frac{3}{a}\rho_m = +\, \widehat{\alpha}\,
\rho_\phi, \quad \frac{d\rho_\phi}{da}+\frac{3}{a}(1+{\rm w}_\phi)
\rho_\phi = -\, \widehat{\alpha}\, \rho_\phi,
\end{equation}
in which case
\begin{equation}
\rho_m = \frac{\rho_0}{a^3}+\frac{\widehat{\alpha}}{a^3} \int
\rho_\phi\, a^3 da, \quad {\rm w}_\phi(a)=-1- \frac{a
(d\rho_\phi/da)}{3\rho_\phi}-\frac{a \widehat{\alpha}}{3}.
\end{equation}
Now, the last term in the expression of ${\rm w}_\phi$ does not
depend on the ratio $\rho_m/\rho_\phi$, but only on the product
$\widehat{\alpha} a$, which can therefore be negligibly small in
the early universe, where $a\ll 1$.

\begin{figure}[ht]
\begin{center}
\hskip-0.3cm
\epsfig{figure=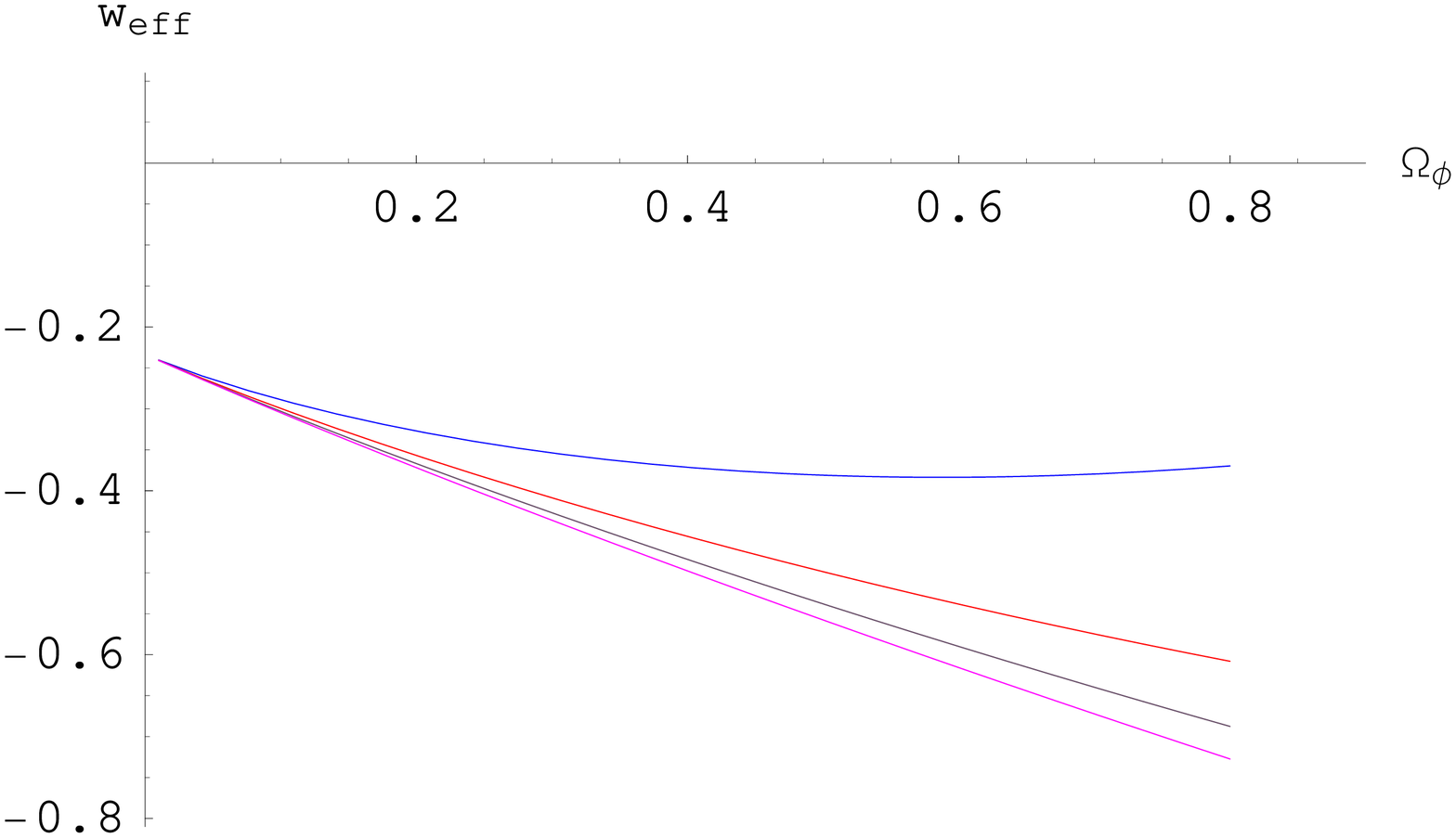,height=2.3in,width=3.3in}
\hskip-0.0cm
\epsfig{figure=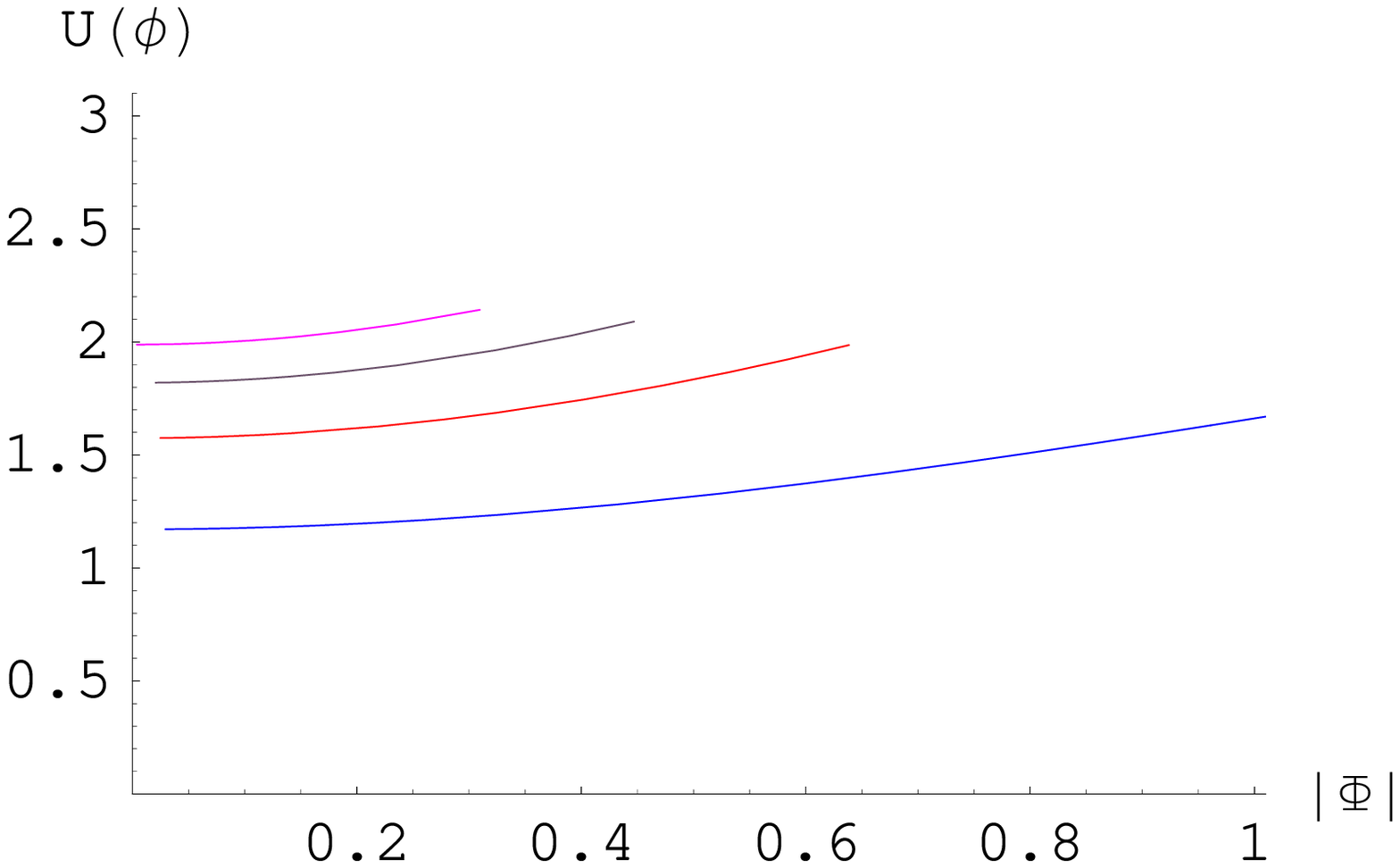,height=2.3in,width=3.2in}
\end{center}
\caption{(Left plot) The effective equation of state ${\rm
w}\Z{\rm eff}$ vs $\Omega\Z{\phi}$ with $\widehat{\beta}=0.8$ and
$n=1, 2, 3, 4$ (top to bottom). (Right plot) The normalised
potential $U(\phi)$ vs $\Phi$, with $\widehat{\beta}=0.8$,
$\Omega_\phi=(0.01,0.8)$ (left to right) and $n=1, 2, 3, 4$
(bottom to top).}\label{Fig7}
\end{figure}

Finally, as one more alternative, let us suppose that
$\beta(\phi)\propto \ln a(\phi)+\beta\Z{0}$. This implies
\begin{equation}
\widehat{\beta}\equiv a \widehat{\alpha} = - a
\frac{d(\kappa\phi)}{da} \frac{d\beta}{d(\kappa\phi)} \equiv {\rm
const}.
\end{equation}
Further, as a phenomenological input, following~\cite{Cai:07a}, we
assume that
\begin{equation}
\rho\Z{\phi}\equiv \frac{3n^2 m\Z{P}^2}{t^2}, \quad t\equiv
\int_0^{a} \frac{da}{Ha}
\end{equation}
where $t>0$. The parameters $\varepsilon$ and ${\rm w}_q$ of the
agegraphic quintessence are now given by
\begin{equation}
\epsilon=-\frac{3}{2}
(1-\Omega\Z{\phi})+\frac{\Omega\Z{\phi}^{3/2}}{n}
-\frac{\widehat{\beta}}{2}(1-\Omega\Z{\phi}), \quad w\Z{\phi} =
-1+\frac{2}{3n}\sqrt{\Omega\Z{\phi}}-\frac{(1-\Omega\Z{\phi})}{\Omega\Z{\phi}}
\frac{\widehat{\beta}}{3},
\end{equation}
where $1>\Omega_\phi>0$. To reconstruct an agegraphic quintessence
potential, we now clearly need an extra input, which is the value
of the coupling $\widehat{\beta}$. With a reasonable choice of the
coupling, say $\widehat{\beta} \lesssim 0.8$, we find that the
shape of the potential $V(\phi)$ is qualitatively similar to that
shown in Fig.~\ref{Fig6}. But we find some other differences (as
compared to the $\widehat{\beta}=0$ case); notably, the universe
can accelerate even if $n\sim {\cal O}(1)$, and the normalised
potential $U(\phi)$ may not vanish at $\Phi=0$ (cf.
Fig.~\ref{Fig7}).

We conclude the Letter with some remarks.\\
The definition (\ref{quantum-CC}), which is, in fact, the central
premise of the agegraghic dark energy proposal, reveals the
possibility that the dark energy density, or gravitational vacuum
energy, at late times is approximated by $\rho\Z{\rm DE}\propto
t\Z{P}^{-2} t\Z{0}^{-2} \sim m\Z{P}^2 H\Z{0}^2$, where $t\Z{0}$ is
mapped to a linear size of the maximum observable patch of the
universe and $H\Z{0}$ is the present value of the Hubble expansion
rate. The form of the agegraphic dark energy as presented in
Eq.~(\ref{quantum-CC}) is problematic if the present universe
consists of only matter and this ``dark energy", possibly for two
reasons. One of which is that one might need a variable $n$ in
order to reconcile the model with the early universe as well as
with dark energy dominance at late times. The other is that the
matter energy density fraction may exhibit some unusual behavior
in the limit $\Omega_q\to 0$. However, both these shortfalls may
be overcome by modifying the ansatz (1), as in Eq. (10), and then
considering a nonzero radiation component in the early universe,
or in the limit $\Omega_q\to 0$. It is interesting to note that
eqn. (\ref{constr-old}) is valid with an arbitrary rescaling in
the definition of agegraphic time $t$, i.e. for both definitions
$\rho_q \propto n^2/t^2$ and $\rho_q\propto n^2/(t+\delta)^2$. The
extra parameter $\delta$ is a kind of mixed blessing, which should
be nonzero in order to satisfy BBN constraints.

For some phenomenologically motivated solutions, like $a\propto
t^{m}$ (where $m=2/3$ during matter dominance and $m>1$ during
dark energy dominance), the matter energy density $\rho_m$ could
be varying as $\rho_m\propto 1/a^3\propto 1/t^2$ and
$\rho_m\propto 1/t^{3m}\ll 1/t^2 \sim \rho_q$, respectively,
during the matter and dark energy dominated epochs. Thus, for a
suitable choice of $n$, the ``agegraphic" dark energy density may
exceed the matter energy density (at late times), leading to a
regime of dark energy dominance.

We have shown that in the case of a non-minimal coupling between
the $q$-field and matter, the model proposed in~\cite{Cai:07a} can
be adjusted to present-day dark energy parameters
$\Omega_{q}\simeq 0.73$ and ${\rm w}_{q}\simeq -1$, by allowing a
relatively large coupling between the $q$-field and (dark) matter.
Although the model does not explain much about the dynamics or the
origin of dark energy, it provides an interesting kinematic
approach to dark energy equation of state by outlining a possible
time growth of dark energy component (at late times). The model
naturally predicts an interesting value for the dark energy
equation of state, which is $-1\le {\rm w}\Z{q}<-1/3$ in the
minimal coupling case. It can be hoped that future cosmological
observations will provide new constraints on this model, via a
more precise measurement of the dark energy equation of state,
which is currently constrained to be $-1.38< {\rm w}_{q} < -0.82$
at zero redshift. The model deserves further investigations,
especially, in the case of a non-minimal interaction between the
$q$-field and (dark) matter.

\medskip
{\bf Note added}: After the first submission of this Letter to the
archive, there have appeared some generalisations of the original
agegraphic dark energy model, including the ${\rm w}$--${\rm
w}^\prime$ phase-space analysis~\cite{Wei:2007z}, the study of
instability of agegraghic dark energy~\cite{Kim-Myung} and
reconstructions of agegraphic quintessence
models~\cite{Wu-Ma-Ling}.

\medskip
{\sl Acknowledgements}: I would like to thank the referee for
making quite useful suggestions, which helped to improve the
overall presentation. This research is supported in part by the
FRST Research Grant No. E5229 (New Zealand) and Elizabeth Ellen
Dalton Research Award (2007). The author acknowledges useful
correspondences with Rong-Gen Cai and Hao Wei.

\medskip

\section*{Appendix A}
\renewcommand{\theequation}{A.\arabic{equation}}
\setcounter{equation}{0}

Here we write the matter Lagrangian ${\cal L}_m$ in a general
form~\cite{Ish:07a}:
\begin{equation}\label{matter-scalar}
{\cal L}_m\equiv {\cal L} (\beta^2(\phi) g_{\mu\nu},\psi_m)=
\sqrt{-g}\, \beta^4(\phi) \sum \rho_i,
\end{equation}
where $\psi\Z{m}$ denotes collectively the matter degrees of
freedom and $\beta(q)$ is a general function of $q$. The radiation
term $\rho_{r}$ ($i=r$) does not contribute to the effective
potential or the Klein-Gordon equation. As a result, the effect of
the coupling $\beta(\phi)$ can be negligibly small during the
epoch where ($\rho\Z{m}\ll \rho\Z{r}$). However, as explained
in~\cite{Ish:07b}, the coupling $\beta(\phi)$ between the
dynamical field $\phi$ and the matter can be relevant especially
in a background where $\rho\Z{m}\gtrsim \rho\Z{r}$ (see, for
example, Refs.~\cite{Damour:1993id}).

Einstein's equations following from Eqs. (\ref{action}) and
(\ref{matter-scalar}) are
\begin{eqnarray}
3 H^2 &=& \kappa^2 \left( \frac{1}{2}\, \dot{\phi}\,^2 + V(\phi)
+\beta^4 \left(\rho_{m}+\rho_{r}\right) \right),\label{nonminimal1}\\
- 2 \dot{H} &=& \kappa^2 \left( \dot{\phi}\,^2  + \beta^4 \left(1+
{\rm w}_m\right) \rho_{m}  +\frac{4}{3} \beta^4 \rho_r\right),
\label{nonminimal2}
\end{eqnarray}
where ${\rm w}_i\equiv p_{i}/\rho_{i}$ and $\rho_{i} \propto
(a\beta)^{-\,3(1+{\rm w}_{i})}$. The equation of motion for $\phi$
is
\begin{equation}
 \ddot{\phi}  +
3H\dot{\phi}= -\frac{dV(\phi)}{d\phi}+\eta Q \beta^4
\rho_{i},\label{nonminimal3}
\end{equation}
and the fluid equation of motion for matter (m) or radiation (r)
is:
\begin{equation}
\dot{\rho}_{i} +3H \rho_{i} (1+ {\rm w}_{i}) =  -\dot{\phi} \eta Q
\beta^4 \rho_{i}, \quad (i=m, r), \label{background1}
\end{equation}
where $\eta\equiv (1-3 {\rm w}_{i})$ and $ Q \equiv \frac{d\ln
\beta(\phi)}{d (\kappa \phi)}$. Eq. (\ref{nonminimal3}) can be
written as
\begin{equation}
\dot{\rho}_{\phi} + 3 H \rho_{\phi} \left(1 + {\rm
w}_{\phi}\right) = \dot{\phi}\eta Q \beta^4
\rho_{m},\label{non-mini4}
\end{equation}
where ${\rm w}_{\phi}\equiv p_{\phi}/\rho_{\phi}$, $\rho_{\phi}
\equiv \frac{1}{2} \dot{\phi}^2 + V(\phi)$ and $p_{\phi} \equiv
\frac{1}{2} \dot{\phi}^2 - V(\phi)$. This last equation along with
(\ref{background1}) guarantees the conservation of total energy:
$\dot{\rho}_{\rm tot}+ 3H (\rho_{\rm tot} + p_{\rm tot})=0$, where
$\rho_{\rm tot}=\rho_{m}+\rho_{r}+ \rho_{\phi}$. Using the
following definitions
\begin{equation}
\varepsilon\equiv \frac{\dot{H}}{H^2},\quad \Omega_{i} \equiv
\kappa^2 \frac{\beta^4 \rho_{i}}{3H^2}, \quad
\Omega_{\phi}=\kappa^2\frac{\rho_{\phi}}{3H^2},
\end{equation}
we arrive at the system of equations~(\ref{EoM1})-(\ref{EoM3}).

\end{document}